\documentclass[reprint, superscriptaddress, onecolumn, aps, pre]{revtex4-2}
\usepackage{amsfonts, amssymb, amsmath}
\usepackage{epstopdf}
\usepackage{epsfig}
\usepackage{hyperref}
\usepackage{color}
\usepackage[utf8]{inputenc}
\usepackage{amsfonts, amssymb, amsmath}
\usepackage{xcolor}
\usepackage{graphics}
\usepackage{float}
\date{September 2021}
\begin{document}

\title{Solving Triad Reduced Interactions for Drift Kinetic Equations}

\author{Evgeny A. Gorbunov}
\affiliation{Coventry University, Coventry CV1 5FB, United Kingdom }
\author{Bogdan Teaca}
\affiliation{Coventry University, Coventry CV1 5FB, United Kingdom }
\affiliation{University of Craiova, 13 A.I. Cuza Street, 200585 Craiova, Romania}

\begin{abstract}
We present the TRIDK code, which is written in \emph{Python}, that solves Triad Reduced Interactions for a Drift Kinetic system of equations. The four-dimensional drift kinetic system of equations captures nonlinear electromagnetic (or electrostatic) interactions for a magnetized plasma at scales larger than the ion gyroradius. In the TRIDK code, the nonlinear interaction are modelled using a compound of nested icosahedra-dodecohedra shells in wave space, which allows to perform computationally efficient simulations over a large span of scales. This reduced approach captures essential nonlinear dynamics, while allowing parallel mixing in velocity space to take place. The intent is to offer a cost efficient way to probe the balance between linear and nonlinear dynamics, and analyse the interplay between linear Landau damping and turbulence.
\end{abstract}

\maketitle

\section{Introduction}
Large scale dynamics in astrophysical plasmas are successfully described by magnetohydrodynamic formalisms \cite{Goldreich:1995p724, Bhattacharjee:1998p313, Zhdankin:2012p1824}. For scales approximately equal or smaller then the ion gyroradius, kinetic descriptions are needed \cite{Schekochihin_2009,Wan:2012p1837}. For magnetized plasma, it is possible to integrate out the fast gyration motion of the particles, making it possible to study collisionless regimes at the scales close to ion gyroradius using gyrokinetics \cite{Howes:2008p1132, Tatsuno:2009p1096}. However, the kinetic description of the plasma at spacial scales larger then gyroradius can be reduced to set of fluid-like equations \cite{zocco2011,Brizard:1992}. These models combine computational effectiveness yet providing accurate description of kinetic effects arising at the aforesaid transition. In astrophysical plasma, where perpendicular characteristic scales (usually a size of a gyroradius) are much smaller compared to parallel (1 AU), use of such reduced kinetic models is beneficial. The important feature of such models is that they are able to capture the transition between kinetic and fluid-like plasmas, making it particularly interesting for turbulence studies \cite{Meyrand2019, meyrand2021}. 

In order to study this fluid-to-kinetic transition, it is possible to derive drift kinetic set of equations for magnetised plasma in a straight magnetic guide field \cite{gorbunov:2021}. These equations were obtained from gyrokinetic system of equations in slab geometry \cite{Howes:2006p1280, Schekochihin_2009, Teaca:SGS2021} by employing a Laguerre decomposition on the system in perpendicular velocity direction \cite{mandell2018}, and retaining only the dominant contributions to the dynamics due to gyration of the particles. This approach is taking only the first two Laguerre moments of the system \cite{gorbunov:2021}. Therefore, the equation obtained represent a four-dimensional system of the equations, with three spatial dimensions and one dimension representing parallel velocity.  The resulting model conserves free energy, and describes the electromagnetic turbulence in the range of scales of $k_\perp \rho_i \approx 1$, since the system takes into account dominant finite-Larmor radius (FLR) effects. The parallel velocity dependence is described via Hermite representation. 

In this paper, we describe the Triad Reduced Interactions Drift Kinetic code developed to solve the four dimensional drift kinetic system of equations that were derived in \cite{gorbunov:2021}. This code was developed in order to provide a simple, computationally inexpensive toy model which allows to probe different physical phenomena, as well as model nonlinear triad interactions - therefore making it applicable to study turbulence. To reduce the load of the computations when modelling nonlinear interactions, spacial scales were discretized using the nested polyhedra model \cite{gurcan2017}, which discretize Fourier space with locally interacting triads, spaced logarithmically, with each wave vector falling into the vertices of dodecahedron or icosahedron. Initial runs are performed, and a few results are provided as example.

\section{Drift Kinetic equations}
The equations implemented numerically are presented here as in \cite{gorbunov:2021}. The system of equations given describe the evolution of the moments of distribution function $h_{s,l}^m(\mathbf{k})$, with parallel dynamics governed by $m = \overline{0,M}$ Hermite moments, and $l={\{0,1\}}$ Laguerre moments. For electromagnetic plasma at scales larger then ion gyroradius $\rho_i$ a drift kinetic limit $k\rho_i \rightarrow 0$ can be applied, thus leading to capability to study the dynamics of only the first two Laguerre moments, with higher moments becoming negligibly small. System of particles of charge $q_s$ with equlibrium density $n_s$ at a background temperature $T_s$ with thermal velocity $v_{Ts} = \sqrt{\frac{2 T_s}{m_s}}$ and background magnetic field $B_0$ is then represented by by $M\times2$ gyrokinetic distribution function moments, and their evolution is governed by drift kinetic equations expressed in Fourier space
\begin{align}\label{eq:Vlasov0}
         \frac{\partial{g}^m_l}{\partial t} = \mathcal{N}_l^m(g) + \mathcal{L}_l^m(h) + C[h_l^m(\mathbf{k})],
\end{align}
with right hand sides expressed as combination of nonlinear, linear and collision terms. Nonlinear behaviour is governed by the interactions between distribution function and gyrokinetic potentials:
\begin{align}\label{eq:nonlinear}
    &\mathcal{N}_0^m(g) = - \frac{1}{B_0}\Bigg[\left\{ \chi^{\phi}+\chi^{B}, g^m_0\right\} + \sqrt{\frac{m+1}{2}}\left\{\chi^{A} ,g_0^{m+1}\right\}+\sqrt{\frac{m}{2}}\left\{\chi^{A} ,g^{m-1}_0\right\}+\left\{ \chi^{B}, g^m_1\right\} \Bigg],\\
    &\mathcal{N}_1^m(g) = - \frac{1}{B_0}\Bigg[\left\{ \chi^{\phi}+\chi^{B}, g^m_1\right\} + \sqrt{\frac{m+1}{2}}\left\{\chi^{A} ,g_1^{m+1}\right\} +\sqrt{\frac{m}{2}}\left\{\chi^{A} ,g^{m-1}_1\right\}+\left\{ \chi^{B}, g^m_0\right\}+2\left\{ \chi^{B}, g^m_1\right\} \Bigg]
\end{align}

Nonlinear terms are represented by a linear combination of Poisson brackets, which in Fourier space are computed via convolution
\begin{equation}\label{eq:poisson}
\{\mathcal{A},\mathcal{B}\} = \frac{1}{2}\sum_{\mathbf{p}+\mathbf{q} =\mathbf{k}}(p_x q_y - p_y q_x)\left(\mathcal{A}_{\mathbf{q}}\mathcal{B}_{\mathbf{p}}-\mathcal{A}_{\mathbf{p}}\mathcal{B}_{\mathbf{q}}\right).
\end{equation}
Linear terms are expressed as following:
\begin{align}
    L_l^m(h) = - i k_z v_{T_s}\left(\sqrt{\frac{m+1}{2}}h^{m+1}_1 + \sqrt{\frac{m}{2}}h^{m-1}_1\right) .
\end{align}
While the form of collision term $C[h_{s,l}^m\mathbf{k}]$ remain arbitrary and can easily be changed, our choice is to have   weak local collisions of frequencies $\nu_\parallel$ in Hermite space and $\nu_\perp$ in Fourier space. The purpose of the collision term is to simply act as a sink of energy at higher Hermite moments and wave numbers:
\begin{equation}\label{eq:collisions}
    C[h_l^m(\mathbf{k})] = -\nu_\parallel m^6 h_l^m(\mathbf{k}) - \nu_\perp k^4 h_l^m(\mathbf{k})\,,
\end{equation}
where for convenience, we use moments of modified gyroradius distribution function, which are linked to gyroradius distribution function through gyrokinetic potentials:
\begin{align}\label{eq:g}
    &g^m_0 = h^m_0 - \frac{q}{T}(\chi^{\phi} +\chi^{B})\delta_{m0} -  \frac{q}{T}\sqrt{\frac{1}{2}}\chi^{A} \delta_{m1}\,,\\
    &g^m_1 = h^m_1 - \frac{q}{T} \chi^{B}\delta_{m0}\,.
\end{align}
Gyrokinetic potentials are given as 
\begin{align}
    \chi^{\phi}(\mathbf{k}) &= \mathcal{J}_{00}\phi(\mathbf{k}) \label{eq:chiphi}\,,\\
    \chi^{B}(\mathbf{k}) &=  \frac{T}{q B_0} \tilde{\mathcal{J}}_{10}B_\parallel(\mathbf{k})\label{eq:chib} \,,\\
    \chi^{A}(\mathbf{k}) &= - v_{T_s} \mathcal{J}_{00} A_\parallel(\mathbf{k})\label{eq:chia}.
\end{align}
 It is important to notice that the equations \eqref{eq:Vlasov0}-\eqref{eq:chia} are capable of capturing the finite Larmor radius (FLR) effects via Bessel function moments 
\begin{align}
    &\mathcal{J}_{00} = e^{-b/2},\label{equation:J round 0}\\
    &\mathcal{\tilde{J}}_{10} = \left[\left(1-e^{-b/2}\right)\frac{2}{b}\right],\label{equation:J round 1}\\
    & b = b_s = \left(\frac{k_\perp v_{Ts}}{\sqrt{2}\Omega_s}\right)^2,\label{eq:b} \\
\end{align}
thus linking to a gyrokinetic model at $k_\perp\rho_i>1$.
Electromagnetic fields are represented by electrostatic potential $\phi(\mathbf{k})=\sum_s  q_s n_s \mathcal{J}_{00}h^0_{s0} \bigg{/} \sum_s\frac{q_s^2 n_s}{T_s}$, 
parallel magnetic field $B_\parallel=-\frac{\beta}{2}\sum_s \frac{n_s T_s}{B_0} \tilde{\mathcal{J}}_{10} \left(h^0_{s0} +h^0_{s1}\right)$
and parallel component of the vector potential $A_\parallel = \frac{\beta}{2 k_\perp^2}\sum_s  q_s n_s v_{T_s} \sqrt{\frac{1}{2}} h^1_{s0}$. It should be noted that the equations \eqref{eq:Vlasov0}-\eqref{eq:chia} are solved for $g(\mathbf{k}_{l,s}^m)$, making it necessary to use the modified distribution function in the definition of electromagnetic fields. This definition is slightly more convoluted, since the electrostatic potential $\phi(\mathbf{k})$ and the parallel component of the magnetic field $B_\parallel(\mathbf{k})$ are coupled via zeroth Laguerre and zeroth Hermite moment of the distribution function. Using the relation \eqref{eq:g} between $h_{ls}^m(\mathbf{k})$ and $g_{ls}^m(\mathbf{k})$, we obtain new definitions for fields:
\begin{align}
    &\phi(\mathbf{k}) = \frac{2 b I_\phi(g)  - \beta c I_{B}(g)}{2 ab + \beta c^2},\label{eq:numerics:phi} \\
    &B_\parallel(\mathbf{k}) = -\beta\frac{a I_{B}(g) + c I_\phi(g)}{2 ab +\beta c^2}.\label{eq:numerics:B}
\end{align}
Here, we introduced a number of parameters for ease of notation,
\begin{align}\label{eq:numerics:coefficients}
        & I_\phi(g) = \sum_s q_s n_s \mathcal{J}_{00} g_{s0}^0, \\
        & I_B(g) = \sum_s \frac{n_s T_s}{B_0} \mathcal{J}_{10} \left(g_{s0}^0 + g_{s1}^0\right),\\
        & a = {\sum_s \frac{q_s^2 n_s}{T_s}}(1 - {\mathcal{J}_{00}^2}), \\
        & b = 1+\beta\sum_s \frac{n_s T_s}{B_0^2}\mathcal{J}_{10}^2, \\
        & c = \sum_s\frac{q_s n_s}{B_0} \mathcal{J}_{10}\mathcal{J}_{00}.
\end{align}
The relation for $A_\parallel$ is also modified, and now expressed via $g_{s0}^1$:
\begin{equation}\label{eq:numerics:A}
    A_\parallel(\mathbf{k}) = \frac{\beta}{2} \frac{\sum_s q_s n_s v_{Ts}\mathcal{J}_{00}\sqrt{\frac{1}{2}}g_{s0}^1}{ k_\perp^2+\frac{\beta}{4}\sum_s \frac{q_s^2 n_s v_{Ts}^2}{T_s}\mathcal{J}_{00}^2}\,.
\end{equation}

The nonlinear terms \eqref{eq:nonlinear} in \eqref{eq:Vlasov0} present a significant computational challenge due to the convolution in \eqref{eq:poisson}. In order to simplify the computation, our choice is to use model with reduced triad interactions, where the exact relation for Poisson brackets \eqref{eq:poisson} is replaced by simpler definition, which is however is able to capture the main nonlinear dynamics arising due to triad interactions. The detailed discussion of the model is given in the next section.

\section{Reduced triad interactions model}

\begin{figure}[H]
    \centering
    \includegraphics[width=0.7\textwidth]{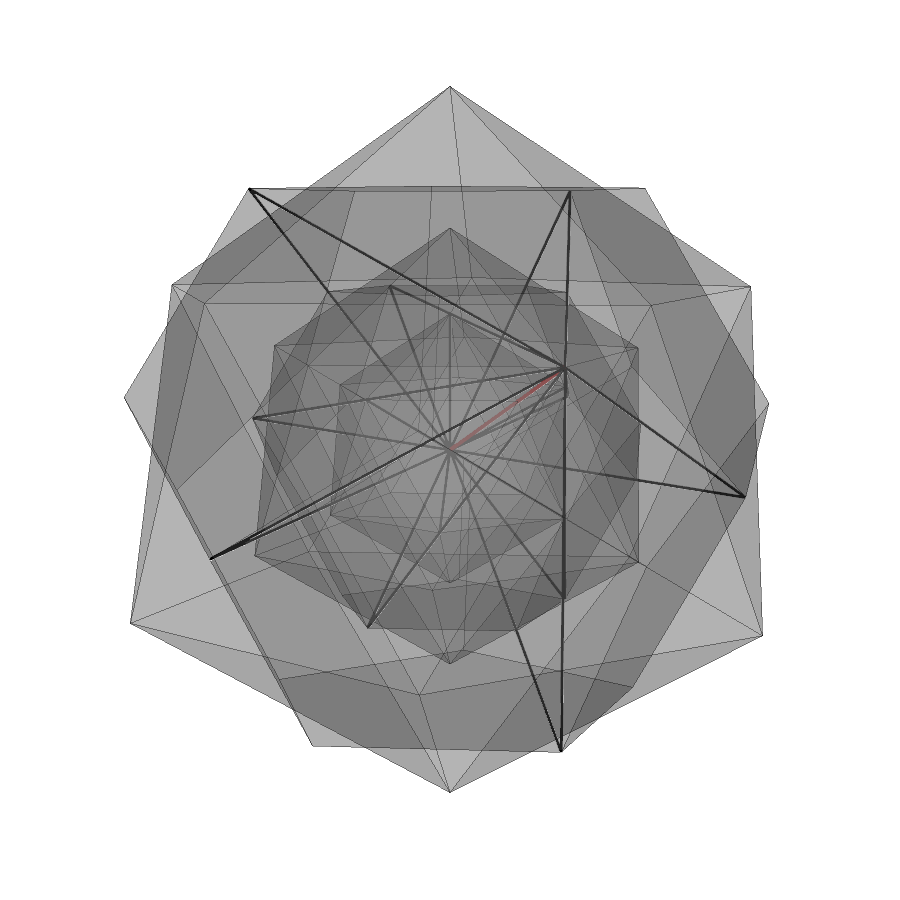}
    \caption{Pairs of wave vectors (black lines) which form a triad with a chosen vector (shown in red) belonging to an icosahedron shell \emph{n}.}
    \label{fig:interacting triads}
\end{figure}

 The nested polyhedra model, which became a model of a choice for TRIDK, is proposed and described in detail in \cite{gurcan2017,gurcan2018}. Here, its  main ideas are described, in order to give the reader better understanding of the model. The wave vector space is discretized with nested, alternating dodecahedra and icosahedra, with their vertices representing wave vector coordinates \textbf{k}. The icosahedron is chosen as a simplest polyhedron ensuring uniform distribution of a wave vectors on a sphere. The dodecahedron was chosen as a dual polyhedron to icosahedron - that is, each vertex of the icosahedron corresponds to one face of dodecahedron, and vice versa. This is the requirement to build triad interactions, which are derived to be local. If we consider a wave vector \textbf{k} falling on the vertex of the shell \emph{n}, then all the possible sets of wave vector pairs $(\mathbf{p},\mathbf{q})$ forming a triad with \textbf{k} belong to the following sets of shells:
\begin{align}
    (\mathbf{p},\mathbf{q})\in\{(n-2,n-1),(n-1,n+1),(n+1,n+2)\}\,.
\end{align}
Each wave vector forms 9 to 15 triads at most, depending the polyhedra type it belongs to. The example of such interacting triads for a single chosen wave vector is shown in fig.\ref{fig:interacting triads}. The interactions due to triads falling out of the boundary of the system (such as they include interactions with \emph{N+1} shell, where \emph{N} is a total number of shells in the system) are considered non-existent.

Since the energy is conserved within a triad, such truncation ensures global conservation of energy for nonlinear interactions. In order to speed up the summation over triads, each wave vector is indexed by unique number \emph{n}. Then, each vector \emph{n} is assigned an indices list $\mathbf{p}_n$ of interacting wave vectors pairs, allowing to compute summation rapidly. The Poisson brackets \eqref{eq:poisson} computed at a node \emph{n} in this case are represented as following:
\begin{align}
    \{\mathcal{A},\mathcal{B}\}_n = \sum_{\{n^{\prime},n^{\prime\prime} \}\in \mathbf{p}_n}(k_{n^\prime,x} k_{n^{\prime\prime},y} - k_{n^{\prime},y} k_{n^{\prime\prime},x})\left(\mathcal{A}_{n^\prime}\mathcal{B}_{n^{\prime\prime}}-\mathcal{A}_{n^{\prime\prime}}\mathcal{B}_{n^{\prime}}\right).
\end{align}
In order to simplify the model even further, the reality condition in Fourier space was used:  
\begin{align}\label{eq:reality}
    \mathcal{A}(\mathbf{k}) = \mathcal{A}^*(-\mathbf{k}).
\end{align} 
this allows us to only consider quantities related to the upper half of the polyhedra $k_z>0$, and restore quantities in full space using the relation \eqref{eq:reality}. When computing \eqref{eq:poisson}, if a triad contains a wave vector which falls onto half-volume $k_z<0$, we use \eqref{eq:reality}, and simply change the sign of the coordinates the wave vector, and complex conjugate all the quantities which are related to it. To do this fast, we for each pair forming a triad with a chosen wave vector we also have a pair of conjugation flags, which tells if the \eqref{eq:reality} should be applied.

The polyhedra are scaled logarithmically \cite{gurcan2017}, and therefore the wave vector coordinates for a shell \emph{n} are defined as $\mathbf{k} = \gamma^n\lambda_{i,d} k_0 \mathbf{\widehat{k}}_l$, with scaling amplitude $\gamma = \sqrt{\frac{1+\sqrt{5}}{2}}$, and $\lambda_i = \sqrt{\frac{\sqrt{5}}{3}}$ for icosahedra and $\lambda_d = 1$ for dodecahedra. The ratio between ocisahedron and dodecahedron, $\lambda_{i,d}$ ensures that two neighbouring polyhedra of the same scale $\gamma^n$ are dual to each other. $\mathbf{\widehat{k}}_l$ is the unit vector describing the positions of the vertex $l$ of the chosen polyhedron. $k_0$ is the scaling factor kept equal to unity for simplicity. 
Due to the chosen discretization, the increase of the system size only leads to linear increase of wave vectors. Therefore, logarithmic scaling allows to cover a large set of scales, without significantly increasing the computational difficulty, with 10 shells covering approximately one decade of k-space.

With all the simplifications given above the nested polyhedra model provides an effective way to compute \eqref{eq:poisson}, while preserving triad interactions and, as a result, conserving free energy. However, logarithmic discretization of the {\emph k}-space makes the system of equations stiff \cite{gurcan2017}. This sets additional constraints on the numerical implementation of such model.
\section{Numerical implementation}

Due to our choice of the discretization of a wavespace, the resulting system of equations \eqref{eq:Vlasov0},\eqref{eq:numerics:phi}-\eqref{eq:numerics:A} becomes stiff. That implies the need to use an implicit A-stable scheme to solve the system of equations. Moreover, stiff systems usually require a Jacobian matrix to be computed in order for the solution to converge. Therefore, the solver which is able to deal with stiff equations fast and effectively must be used. Since the model was originally designed as a quick, 'table-top' tool for probing different physical effects, the code was written purely in \emph{Python}, as it is allows the convenient and fast post-processing of the numerical results. For this work, several options were tried in order to find the most suitable and fast solver:
\begin{enumerate}
    \item \emph{ZVODE} integrator \cite{ZVODE} used in \emph{Python}, and accessed via \emph{SciPy's} \emph{integrate.ode} interface class. This integrator was originally used in \cite{gurcan2017,gurcan2018}. While it is effective for the systems of small sizes, it does not support computations with sparse jacobian matrices, leading to an extremely inefficient consumption of memory scaling as  $O(n^2)$.
    \item Complex Rosenbrock Scheme (CROS) proposed in \cite{Alshin2006}. Promising due to it beautiful simplicity, it has been able to effectively solve linear equations. Unfortunately, it has shown not to be able to provide required accuracy with a reasonably large time step for nonlinear simulations, hence being ineffective for our needs.
    \item  \emph{SciPy}'s \emph{integrate.BDF} class, which is able to work with sparse matrices, having an adaptive time-stepping algorithm, was found to be the best one for the problem of interest. However, it was noticed that for large jacobian matrices, the linear algebraic system \emph{SuperLU} solver used in the class by default becomes slow and inefficient. Therefore, we decided to replace it with \emph{scikit-umfpack}  linear solver package, which gave the speedup of more then 400 in some cases.
\end{enumerate}

\begin{figure}[H]
    \centering
    \includegraphics[width=0.5\textwidth]{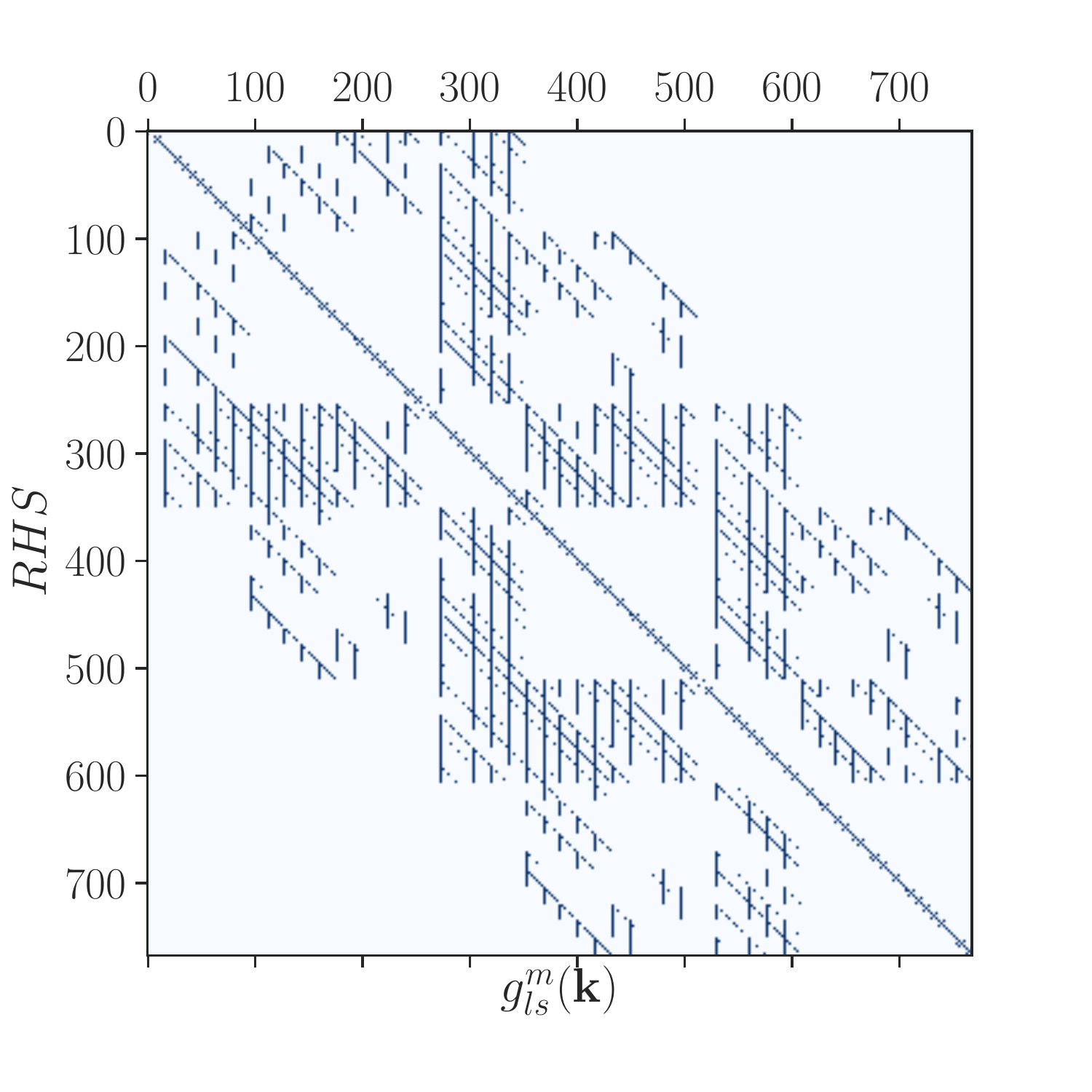}
    \caption{Sparsity structure of the Jacobian $\mathcal{G}_g$. Diagonal elements are due to the collisions contributions to the system. The small dots right above and below the diagonal are due to linear term. The stems and diagonal lines far from the main diagonal are due to nonlinear term contribution.}
    \label{fig:sparsity structure}
\end{figure}

The stiff equations are require the computation of the Jacobian matrix to be solved. For an average TRIDK electromagnetic simulation of size of 30 shells, 50 Hermite moments and 2 species of particles it would require to compute $(48000 \times 48000)$ Jacobian matrix (flattened to be 2D matrix). Fortunately, since one of the core concepts of shell model is that only the local interactions are preserved, the Jacobian matrix remains sparse, significantly simplifying the computations. the Jacobian is being computed in three parts, each related to either nonlinear, linear or collision terms:
\begin{align}\label{eq:jacobian}
    \mathcal{G}_g = \mathcal{G}^{nonlinear}_g+\mathcal{G}^{linear}_g+\mathcal{G}^{collision}_g\,.
\end{align}
The simplest to compute are Jacobian of a linear term $G^{linear}_g$, with each component
\begin{align}
    \mathcal{G}^{linear}_g(\mathbf{k},m,l,s,\mathbf{p},q,j,o) &=\frac{\partial\mathcal{L}^m_l(\mathbf{k})}{\partial g^l_m(\mathbf{p})} =
    \frac{\partial\mathcal{L}^m_l(\mathbf{k})}{\partial h^l_m(\mathbf{p})} \frac{\partial h^l_m(\mathbf{p})}{\partial g^l_m(\mathbf{p})} \nonumber\\
    &= - i k_z\left(\sqrt{\frac{m+1}{2}}\delta_{m+1,q}+\sqrt{\frac{m}{2}}\delta_{m-1,q}\right)\delta_{lj}\delta_{so}\delta_{\mathbf{k}\mathbf{p}}\frac{\partial h^l_m(\mathbf{p})}{\partial g^l_m(\mathbf{p})}\,,
\end{align}
and collision term, with each component
\begin{align}
    \mathcal{G}^{collision}_g(\mathbf{k},m,l,s,\mathbf{p},q,j,o) &=\frac{\partial C[h_l^m(\mathbf{k})]}{\partial g^l_m(\mathbf{p})} =
    \frac{\partial C[h_l^m(\mathbf{k})]}{\partial h^l_m(\mathbf{p})} \frac{\partial h^l_m(\mathbf{p})}{\partial g^l_m(\mathbf{p})} \nonumber\\
    &= -\left(\nu_\parallel m + \nu_\perp k^2 \right)\delta_{mq}\delta_{\mathbf{k}\mathbf{p}}\delta_{lj}\delta_{so}\frac{\partial h^l_m(\mathbf{p})}{\partial g^l_m(\mathbf{p})}\,.
\end{align}
if we introduce the notation  
\begin{align}
    &\mathcal{G}_h^{linear}=\frac{\partial\mathcal{L}^m_l(\mathbf{k})}{\partial h^l_m(\mathbf{p})} = - i k_z\left(\sqrt{\frac{m+1}{2}}\delta_{m+1,q}+\sqrt{\frac{m}{2}}\delta_{m-1,q}\right)\delta_{lj}\delta_{so}\delta_{\mathbf{k}\mathbf{p}},\\
    &\mathcal{G}_h^{collision}=\frac{\partial C[h_l^m(\mathbf{k})]}{\partial h^l_m(\mathbf{p})} = -\left(\nu_\parallel m + \nu_\perp k^2 \right)\delta_{mq}\delta_{\mathbf{k}\mathbf{p}}\delta_{lj}\delta_{so},\\
    &\mathcal{H}_g=\frac{\partial h^l_m(\mathbf{p})}{\partial g^l_m(\mathbf{p})},
\end{align}
then the both linear and collision Jacobians are computed via matrix product
\begin{align}
    &\mathcal{G}_g^{linear}=\mathcal{G}_h^{linear}\cdot \mathcal{H}_g,\\
    &\mathcal{G}_g^{collision}=\mathcal{G}_h^{collision}\cdot\mathcal{H}_g.
\end{align}
Jacobian of the nonlinear term is way harder to compute, since it includes several Poisson brackets \eqref{eq:poisson}. In the TRIDK code, we compute contribution of each Poisson bracket to Jacobian separately, and then sum them together. For example, the contribution to the Jacobian from term $\{\chi^\phi+\chi^B,g_0^m\}$ is computed as following:
\begin{align}\label{nonlinearJac}
    &\frac{\partial}{\partial g^q_j(\mathbf{u})} \{\chi^\phi+\chi^B,g_0^m\} = \frac{1}{2}\frac{\partial}{\partial g^q_j(\mathbf{u})}\sum_{\mathbf{p}+\mathbf{q}+\mathbf{k} = 0}(p_x q_y - p_y q_x)\bigg[\left(\chi^{\phi*}(\mathbf{q})+\chi^{B*}(\mathbf{q})\right)g_0^{m*}(\mathbf{p}) -
    \left(\chi^{\phi*}(\mathbf{p})+\chi^{B*}(\mathbf{p})\right)g_0^{m*}(\mathbf{q})\bigg]\,.
\end{align}
Further derivation is not shown here for simplicity, but is performed in the same manner as for linear and collision Jacobians. 
The sparsity structure of the Jacobian for the system of size  of 6 shells (48 wavevectors), 4 Hermite moments and 2 species is shown in fig.\ref{fig:sparsity structure}.

\section{Numerical results}
In order to verify the code, several numerical tests were performed. To match with \cite{Teaca:SGS2021}, all the test were performed for electron-proton plasma, with mass ratio between ions and electrons $m_i/m_e = 1836$, plasma beta was chosen to be $\beta = 1$, temperature of both species was set to $T_e = T_i = 1 $.
\subsection{Free energy conservation}
\begin{figure}[H]
    \centering
    \includegraphics[width=0.5\textwidth]{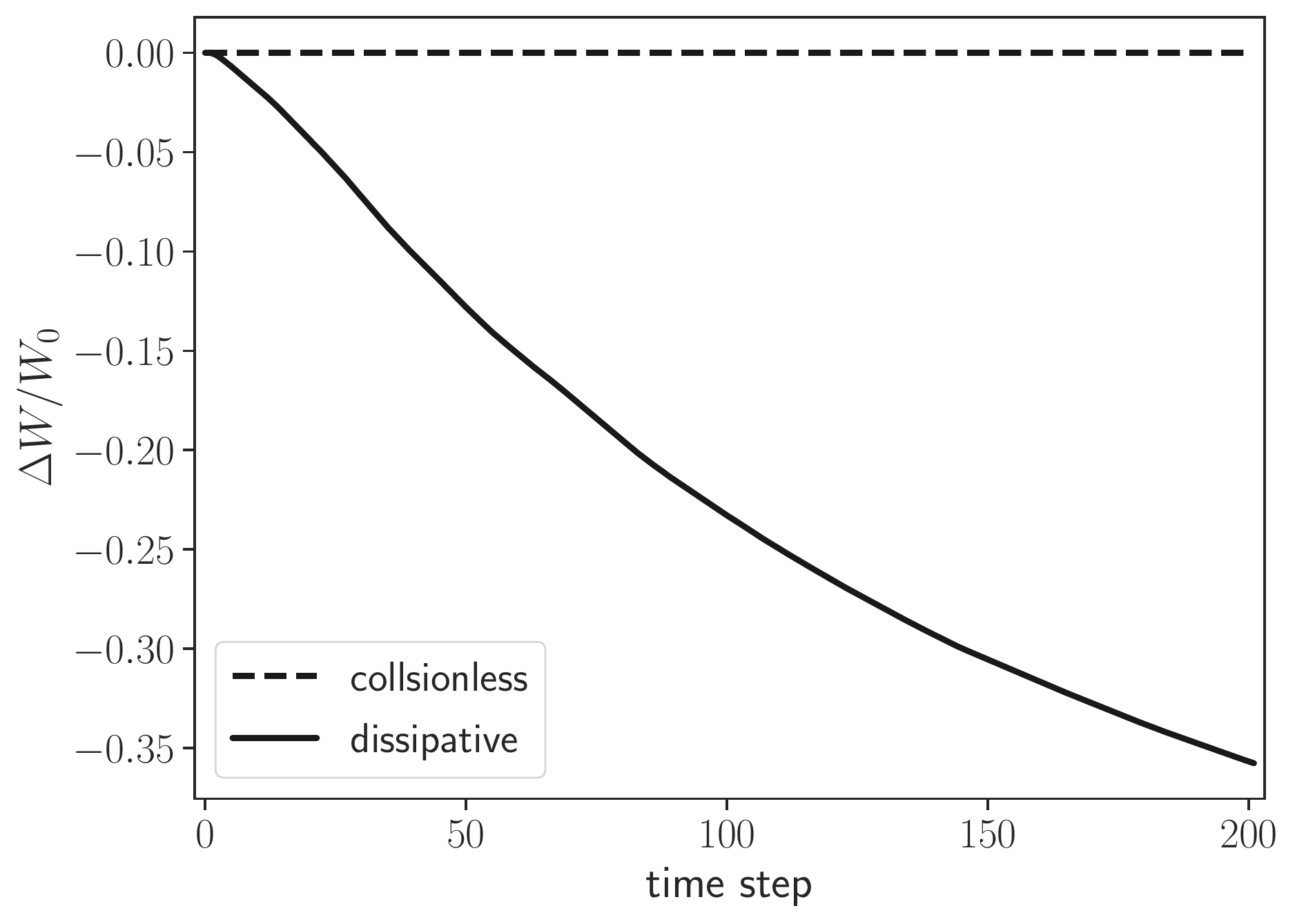}
    \caption{Relative change of energy  $\Delta W/W_0$ 20 shells, 20 Hermite moments. Dashed line demonstrate collisionless simulation. Solid lines show simulation with dissipation. Collision frequencies for that case were taken to be $\nu_\parallel = 10^{-6}$ and $\nu_\perp = 10^{-8}$ . }
    \label{fig:energy conservation}
\end{figure}

For the first test the system was not subjected to collisions in order to check the conservation of free energy,
\begin{align}
    &W(t) = \sum_s  W_s(t),\\
    &W_s(t) = \pi^{5/2} T_s \sum_{\mathbf{k},m,l} \Re(g^m_{l,s}(\mathbf{k}) h^{m*}_{l,s}(\mathbf{k})).
\end{align}
The numerical scheme used have proved to be successful in preserving free energy for the system of different sizes, with relative change of energy $\Delta W/W_0 = \frac{W(t)-W(0)}{W(0)}$ not exceeding the $10^{-6}$ amplitude for 200 time steps. This collisionless cases were then compared to runs with dissipation.  The results of the simulations of size of 20 Hermite moments and 20 shells are shown on fig.\ref{fig:energy conservation}. 

\subsection{Linear run}
\begin{figure}[H]
    \centering
    \includegraphics[width=0.5\textwidth]{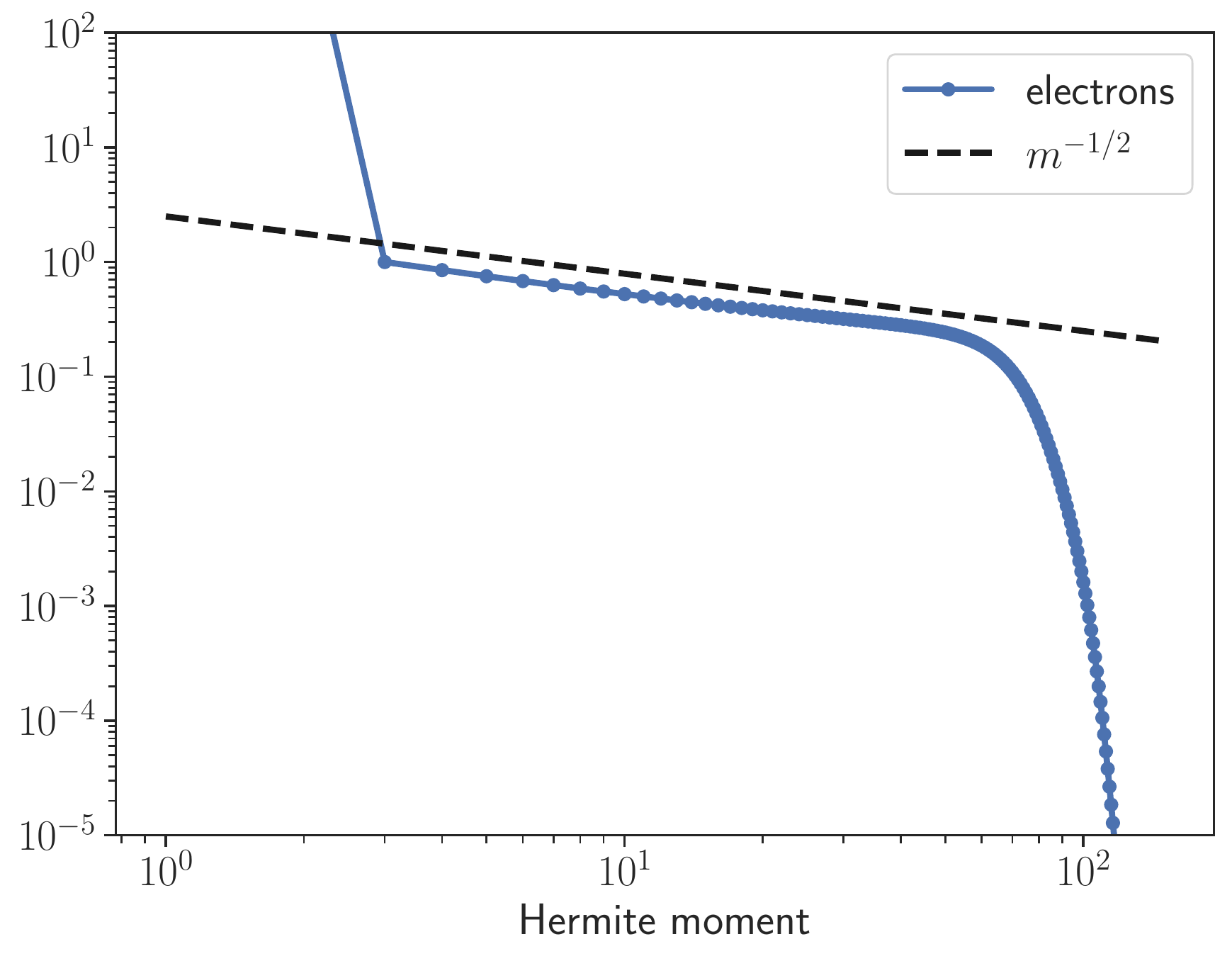}
    \caption{Free energy spectra for linear runs, electrons. Collision frequencies for that case were taken to be $\nu_\parallel = 10^{-10}$ and $\nu_\perp = 0$ . }
    \label{fig:linear run}
\end{figure}

Another test was the linear run performed with system of size 150 Hermite moments and 6 shells. Due to the fact that ion's dynamics are way slower compared to the electrons (due to the high mass ratio), we look only at the electrons in order not to keep the test runs short. The displayed data was normalised by the value of free energy at third Hermite moment, and then averaged over time over a small time window. This has been done in order to obtain smooth spectrum, while taking into account that the energy is being dissipated from the system. The resulting free energy spectra for electron species, is presented in fig.\ref{fig:linear run}. The observed power law is the same as predicted in \cite{zocco2011} from simple theoretical consideration.

\subsection{Nonlinear run}
\begin{figure}[H]
    \centering
    \includegraphics[width=0.5\textwidth]{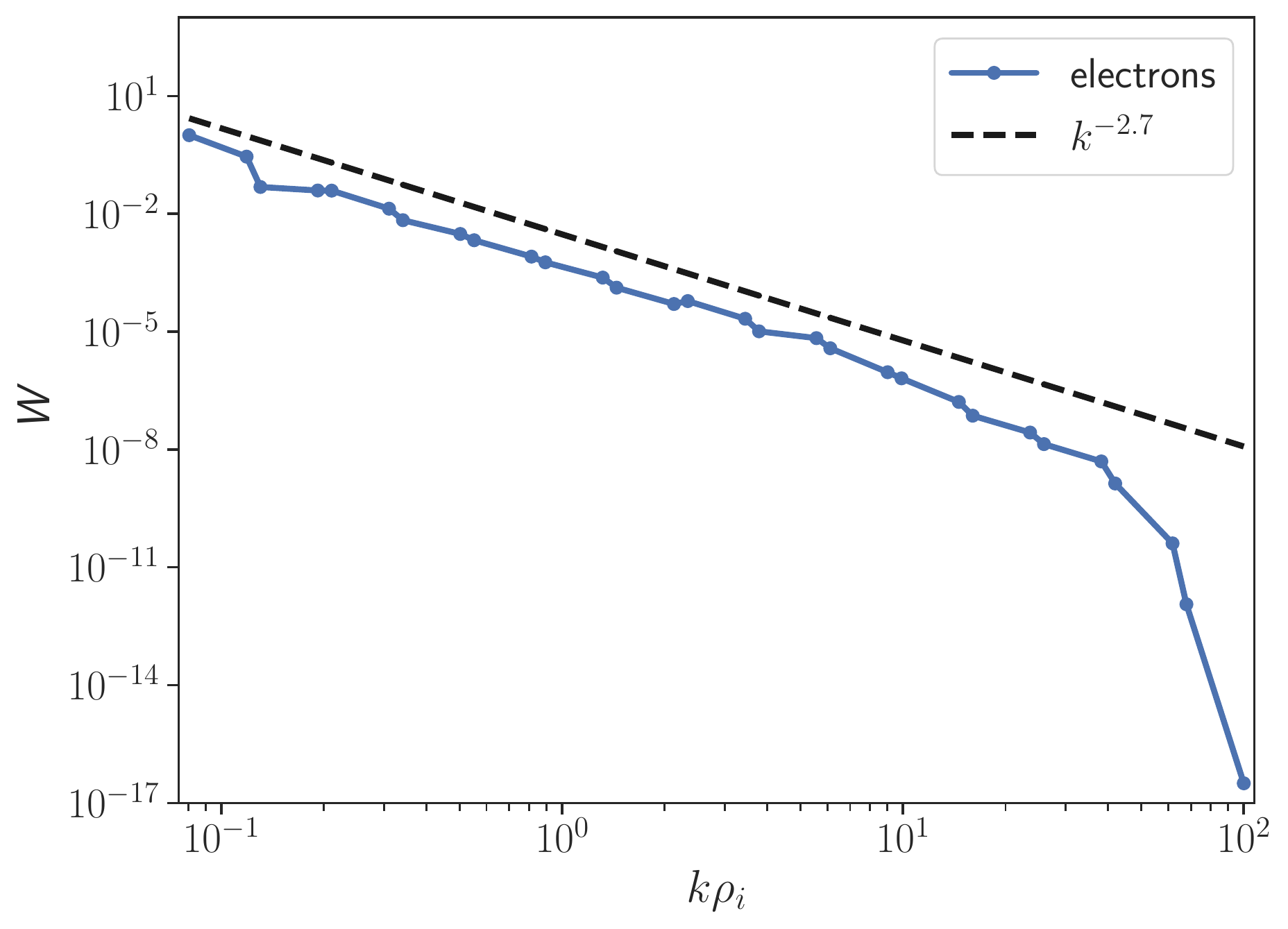}
    \caption{Free energy spectra for nonlinear run. The results are shown for electrons. Collision frequencies for that case were taken to be $\nu_\parallel = 10^{-6}$ and $\nu_\perp = 10^{-6}$ . }
    \label{fig:nonlinear run}
\end{figure}

Initial nonlinear runs was performed as well. The run was performed for 30 shells and 50 Hermite moments. For the same numerical time considerations as for the linear run, only electron spectra is analysed. The data is normalised by the value of free energy at $k \rho_i = 1$. The result of the is shown in fig.\ref{fig:nonlinear run}, demonstrating the power law of $\approx k^{-2.7}$, a spectrum observed for solar wind at sub-ion scales \cite{Kiyani:2015}. This result is encouraging us to continue with the further developing of the code and obtain a speed up for typical runs.
\section{Conclusions}
TRIDK is a code which applies nested polyhedra model \cite{gurcan2017,gurcan2018} to the drift kinetic system \eqref{eq:Vlasov0}-\eqref{eq:numerics:A}. While the resulting system is stiff, TRIDK still provides a good toy model for probing different physical phenomena in electromagnetic plasma turbulence, at scales where fluid turbulence connects to a kinetic one. Numerical test have proved the validity of the model, providing the conservation of free energy and predicting a correct power law in Hermite space for linear simulations. The code has also showed that it is possible to study nonlinear cascade as well.  

Since the code initially was developed as a fast tool for 'table-top' simulations (i.e., for a laptop), \emph{Python} was chosen as a programming language. However, due to the stiffness  of the system the choice of the language is not optimal due to parallelization constraints. In order to provide faster computational speed, TRIDK will be transferred to \emph{Julia} programming language, where it will benefit from compiled language speed up and GPU computations. We consider this step to be needed before TRIDK can be used for regular scientific use. The current presentation is made to further discussions on the topic of using triad reduced nonlinear interactions for the study of kinetic plasmas. 

\end{document}